\documentclass[runningheads]{llncs}
\usepackage{graphicx}
\usepackage{listings}
\usepackage[toc,page, title]{appendix}
\usepackage{listings, xcolor}

\definecolor{verylightgray}{rgb}{.97,.97,.97}

\lstdefinelanguage{Solidity}{
	keywords=[1]{anonymous, assembly, assert, balance, break, call, callcode, case, catch, class, constant, continue, contract, debugger, default, delegatecall, delete, do, else, emit, event, export, external, false, finally, for, function, gas, if, implements, import, in, indexed, instanceof, interface, internal, is, length, library, log0, log1, log2, log3, log4, memory, modifier, new, payable, pragma, private, protected, public, pure, push, require, return, returns, revert, selfdestruct, send, storage, struct, suicide, super, switch, then, this, throw, transfer, true, try, typeof, using, value, view, while, with, addmod, ecrecover, keccak256, mulmod, ripemd160, sha256, sha3}, % generic keywords including crypto operations
	keywordstyle=[1]\color{blue}\bfseries,
	keywords=[2]{address, bool, byte, bytes, bytes1, bytes2, bytes3, bytes4, bytes5, bytes6, bytes7, bytes8, bytes9, bytes10, bytes11, bytes12, bytes13, bytes14, bytes15, bytes16, bytes17, bytes18, bytes19, bytes20, bytes21, bytes22, bytes23, bytes24, bytes25, bytes26, bytes27, bytes28, bytes29, bytes30, bytes31, bytes32, enum, int, int8, int16, int24, int32, int40, int48, int56, int64, int72, int80, int88, int96, int104, int112, int120, int128, int136, int144, int152, int160, int168, int176, int184, int192, int200, int208, int216, int224, int232, int240, int248, int256, mapping, string, uint, uint8, uint16, uint24, uint32, uint40, uint48, uint56, uint64, uint72, uint80, uint88, uint96, uint104, uint112, uint120, uint128, uint136, uint144, uint152, uint160, uint168, uint176, uint184, uint192, uint200, uint208, uint216, uint224, uint232, uint240, uint248, uint256, var, void, ether, finney, szabo, wei, days, hours, minutes, seconds, weeks, years},	% types; money and time units
	keywordstyle=[2]\color{teal}\bfseries,
	keywords=[3]{block, blockhash, coinbase, difficulty, gaslimit, number, timestamp, msg, data, gas, sender, sig, value, now, tx, gasprice, origin},	% environment variables
	keywordstyle=[3]\color{violet}\bfseries,
	identifierstyle=\color{black},
	sensitive=false,
	comment=[l]{//},
	morecomment=[s]{/*}{*/},
	commentstyle=\color{gray}\ttfamily,
	stringstyle=\color{red}\ttfamily,
	morestring=[b]',
	morestring=[b]"
}

\usepackage{tabularx}
\usepackage{array}
\usepackage{multirow}
\usepackage{amsmath}
\usepackage{amssymb}
\usepackage{why3lang}
\usepackage{wrapfig}
\usepackage{booktabs}

\begin{document}
\title{Deductive Proof of Industrial Smart Contracts Using Why3}
\author{Zeinab Neha\"{\i}\inst{1,2} \and
Fran\c cois Bobot\inst{2}}
\authorrunning{Z. Neha\"{\i} \& F. Bobot}
\institute{Universit\'{e} Paris Diderot, Paris, France \and
CEA LIST, Palaiseau, France \\
\email{zeinab.nehai@univ-paris-diderot.fr} \\
\email{\{zeinab.nehai, francois.bobot\}@cea.fr}}
\maketitle              % typeset the header of the contribution
\begin{abstract}
%A bug or error is a common problem that any software or computer program may encounter. It can occur from badly writing the program, a typing error or bad memory management. However, errors can become a significant issue if the unsafe program is used for critical systems. Therefore, formal methods for these kinds of systems are greatly required. 
In this paper, we use a formal language that performs deductive verification on industrial smart contracts, which are self-executing digital programs. Because smart contracts manipulate cryptocurrency and transaction information, if a bug occurs in such programs, serious consequences can happen, such as a loss of money. The aim of this paper is to show that a language dedicated to deductive verification, called \textit{Why3}, can be a suitable language to write correct and proven contracts. We first encode existing contracts into the \textit{Why3} program; next, we formulate specifications to be proved as the absence of RunTime Error and functional properties, then we verify the behaviour of the program using the \textit{Why3} system. Finally, we compile the \textit{Why3} contracts to the Ethereum Virtual Machine (EVM).  Moreover, our approach estimates the cost of gas, which is a unit that measures the amount of computational effort during a transaction.
\end{abstract}
%Moreover, we propose a formalized library that encodes features of Solidity as primitive functions or specific types.
%Moreover, we give a set of generic mathematical statements that allows verifying functional properties suited to any type of smart contracts holding cryptocurrency, showing that \textit{Why3} can be a suitable language to write smart contracts.
%

\section{Introduction}
\paragraph{}Smart Contracts~\cite{wood2014ethereum} are sequential and executable programs that run on Blockchains~\cite{nakamoto2008bitcoin}. They permit trusted transactions and agreements to be carried out among parties without the need for a central authority while keeping transactions traceable, transparent, and irreversible. 
%A contract can define any set of rules represented in its programming language, thus enabling the implementation of decentralized applications. 
These contracts are increasingly confronted with various attacks exploiting their execution vulnerabilities. Attacks lead to significant malicious scenarios, such as the infamous \textit{The DAO} attack~\cite{atzei2017survey}, resulting in a loss of $\sim$\$60M. In this paper, we use formal methods on smart contracts from an existing Blockchain application. Our motivation is to ensure safe and correct contracts, avoiding the presence of computer bugs, by using a deductive verification language able to write, verify and compile such programs. 
%Specifically, the desired behaviors of the program are defined by pre-conditions that characterize a condition that must be true before the beginning of a program, and post-conditions that say what is true at the end of the program. 
The chosen language is an automated tool called \textit{Why3}~\cite{filliatre2013why3}, which is a complete tool to perform deductive program verification, based on Hoare logic. A first approach using \textit{Why3} on solidity contracts (the Ethereum smart contracts language) has already been undertaken~\cite{SolandWhy}. The author uses \textit{Why3} to formally verify \textit{Solidity} contracts based on code annotation. Unfortunately, that work remained at the prototype level. We describe our research approach through a use case that has already been the subject of previous work, namely the Blockchain Energy Market Place (BEMP) application~\cite{nehai}. In summary, the contributions of this paper are as follows: 
%Since its creation, \textit{Solidity} has been in constant evolution and development in order to create a secure environment against potential attacks. As a result, \textit{Atzei et al.} in~\cite{atzei2017survey} have established a summary of common programming pitfalls and identified two major causes of vulnerabilities: problems in the \textit{Solidity} language and poor documentation of weaknesses. Moreover, they give a taxonomy of vulnerabilities in Ethereum smart contracts.
%It has a logic language in which we express the specifications and check the proofs, and a programming language in which we define the program. 

\begin{enumerate}
\item Showing the adaptability of \textit{Why3} as a formal language for writing, checking and compiling smart contracts. 
%This approach gives us a correct-by-construction programs based on a case study.
\item Comparing existing smart contracts, written in \textit{Solidity}~\cite{buterin2014next}, and the same existing contracts written in \textit{Why3}. 
%Showing that \textit{Why3} can properly encode \textit{Solidity} language.
\item Detailing a formal and verified \textit{Trading} contract, an example of a more complicated contract than the majority of existing \textit{Solidity} contracts.
\item Providing a way to prove the quantity of \textit{gas} (fraction of an Ethereum token needed for each transaction) used by a smart contract.
\end{enumerate}
The paper is organized as follows. Section 2 describes the approach from a theoretical and formal point of view by explaining the choices made in the study, and section 3 is the proof-of-concept of compiling \textit{Why3} contracts. A state-of-the-art review of existing work concerning the formal verification of smart contracts is described in section 4. Finally, section 5 summarizes conclusions.

\section{A New Approach to Verifying Smart Contracts Using Why3}
\subsection{Background of the study}
\begin{wrapfigure}[13]{r}{0.50\textwidth}
    \centering
    \includegraphics[width=0.50\textwidth]{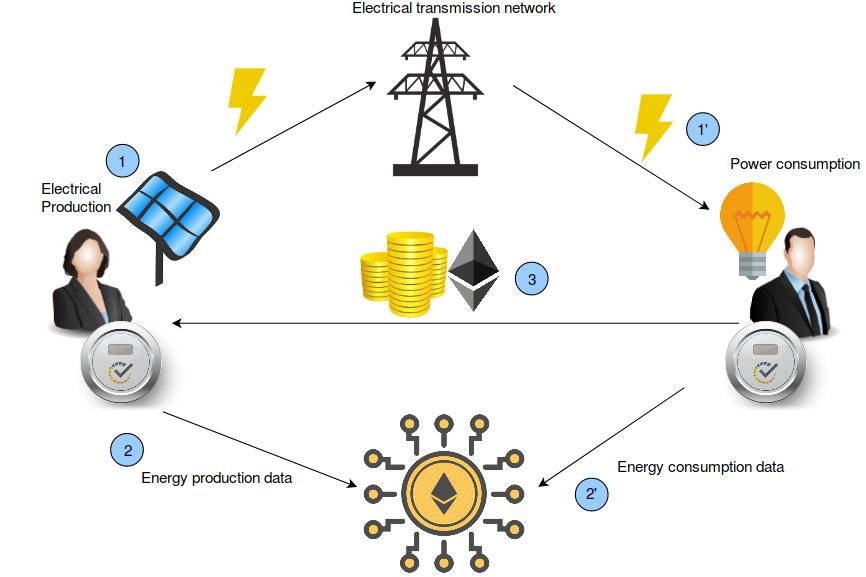}
    \caption{BEMP Process}
    \label{fig:modelingBEMP}
\end{wrapfigure}

\paragraph{Deductive approach \& Why3 tool.} A previous work aimed to verify smart contracts using an abstraction method, model-checking~\cite{nehai}. 
%The approach was to model the operation of the application and to formulate properties in temporal logic in order to verify that the model satisfies the properties. The three-fold model consists of the Kernel layer that models the Ethereum blockchain behavior, the application layer that models functions of the smart contracts, and finally the environment layer that models the oracle framework. 
Despite interesting results from this modelling method, the approach to property
verification was not satisfactory. Indeed, it is well-known that model-checking
confronts us either with limitation on combinatorial explosion, or limitation
with invariant generation. Thus, proving properties involving a large number of
states was impossible to achieve because of these limitations. This conclusion
led us to consider applying another formal methods technique, deductive
verification, which has the advantage of being less dependent on the size of the
state space. In this approach, the user is asked to write the invariants. We
chose the automated \textit{Why3} tool~\cite{filliatre2013why3} as our platform
for deductive verification. It provides a rich language for specification and
programming, called \textit{WhyML}, and relies on well-known external theorem
provers such as Alt-ergo~\cite{altergo}, Z3~\cite{z3smt}, and
CVC4~\cite{BCD+11}.
% In our case study, we chose Alt-ergo.
\textit{Why3} comes
with a standard library\footnote{http://why3.lri.fr/} of logical theories and
programming data structures. The logic of \textit{Why3} is a first-order logic
with polymorphic types and several extensions: recursive definitions, algebraic
data types and inductive predicates.
\paragraph{Case study: Blockchain Energy Market Place.} We have applied our approach to a case study provided by industry~\cite{nehai}. It is an Ethereum Blockchain application (BEMP) based on \textit{Solidity} smart contracts language. 
%Furthermore, Why3 itself is a suitable language designed to aid in the analysis of computer programs, where the source code of a \textit{Solidity} program is translated into a form more suitable for code-improving transformations. 
Briefly, this Blockchain application makes it possible to manage energy exchanges in a peer-to-peer way among the inhabitants of a district as shown in Figure~\ref{fig:modelingBEMP}. The figure illustrates (1) \& (1') energy production (Alice) and energy consumption (Bob). (2) \& (2') Smart meters provide production/consumption data to Ethereum blockchain. (3) Bob pays Alice in \textit{ether} (Ethereum's cryptocurrency) for his energy consumption. For more details about the application, please refer to~\cite{nehai}.
\paragraph{} In our initial work, we applied our method on a simplified version of the application, that is, a one-to-one exchange (1 producer and 1 consumer), with a fixed price for each kilowatt-hour. This first test allowed us to identify and prove RTE properties. The simplicity of the unidirectional exchange model did not allow the definition of complex functional properties to show the importance and utility of the \textit{Why3} tool. In a second step, we extended the application under study to an indefinite number of users, and then enriched our specifications. The use of \textit{Why3} is quite suitable for this order of magnitude. In this second version, we have a set of consumers and producers willing to buy or to sell energy. Accordingly, we introduced a simple trading algorithm that matches producers with consumers. In addition to transferring \textit{ether}, users transfer crypto-Kilowatthours to reward consumers consuming locally produced energy. Hence, the system needs to formulate and prove predicates and properties of functions handling various data other than cryptocurrency. For a first trading approach, we adopted, to our case study, an order book matching algorithm~\cite{domowitz1993taxonomy}.

%Moreover, we chose to switch from the \textit{Solidity} program to the \textit{Why3} because \textit{Solidity} changes very frequently to address attacks; therefore its semantics is not clear enough to directly prove the source code. We believe \textit{Why3} could be a language for writing smart contracts while proving their correctness and absence of bugs.
\subsection{Why3 features intended for Smart Contracts }
\subsubsection{Library modelling.} \textit{Solidity} is an imperative
object-oriented programming language, characterized by static
typing\footnote{Ethereum foundation: Solidity, the contract-oriented programming
  language. https://github.com/ethereum/solidity}. It provides several
elementary types that can be combined to form complex types such as booleans,
signed, unsigned, and fixed-width integers, settings, and domain-specific types
like addresses. Moreover, the address type has primitive functions able to
transfer \textit{ether} (\verb|send()|, \verb|transfer()|) or manipulate
cryptocurrency balances (\verb|.balance|). \textit{Solidity} contains elements
that are not part of the \textit{Why3} language. One could model these as
additional types or primitive features. Examples of such types are
\verb|uint256| and \verb|address|. For machine integers, we use the range
feature of Why3: \lstinline[language = Why3, basicstyle=\fontsize{7}{9}\tt]
! type uint256 = <range 0 0x7FFFFFFFFFFFFFFFFFFFFFFFFFFFFF... >! because it
exactly represents the set of values we want to represent. Moreover why3 checks that
the constants written by the user of this types are inside the bounds and
converts in specifications automatically range types to the mathematical
integers, e.g., \verb|int| type. Indeed it is a lot more natural and clearer to
express specification with mathematical integers, for example with wrap-around semantic \lstinline[language
= Why3, basicstyle=\fontsize{7}{9}\tt] ! account = old account - transfer !
doesn't express that the account lose money (if the account was empty it could
now have the maximum quantity of money).
% Then we introduce values that allow the conversion from \verb|uint256| to \verb|int| (and conversely) in the program.
% \begin{lstlisting}[language = Why3]
%     val uint256_to_int (n : uint256) : int
%         requires {"expl:integer overflow" in_bounds (to_int n)}
%         ensures {result = to_int n}
% \end{lstlisting}

Based on the same reasoning, we have modelled the type \verb|Int160|,
\verb|Uint160| (which characterizes type \verb|uint| in \textit{Solidity}). We
also model the \verb|address| type and its members. We choose to encode the
private storage (\verb|balance|) by a Hashtable having as a key value an
address, and the associated value a \verb|uint256| value. The current value of
the balance of addresses would be \verb|balance[address]|. In addition, the
\verb|send| function is translated by a \verb|val| function, which performs
operations on the \verb|balance| hashtable. Moreover, we model primitive
features such as the \verb|modifier| function, whose role is to restrict access
to a function; it can be used to model the states and guard against incorrect
usage of the contract. In \textit{Why3} this feature would be an exception to be
raised if the condition is not respected, or a precondition to satisfy. We will
explain it in more details with an example later. Finally, we give a model of
\textit{gas}, in order to specify the maximum amount of \textit{gas} needed in
any case. We introduce a new type: \verb|type gas = int|. The quantity of
\textit{gas} is modelled as a mathematical integer because it is never
manipulated directly by the program. This part is detailed later.

It is important to note that the purpose of our work is not to achieve a
complete encoding of \textit{Solidity}. The interest is rather to rely on the
case study in our possession (which turns out to be written in
\textit{Solidity}), and from its contracts, we build our own \textit{Why3}
contracts. Therefore, throughout the article, we have chosen to encode only
\textit{Solidity} features encountered through our case study. Consequently,
notions like \verb|revert| or \verb|delegatecall| are not treated. Conversely,
we introduce additional types such as \verb|order| and \verb|order_trading|,
which are specific to the BEMP application. The \verb|order| type is a record
that contains \verb|orderAddress| which can be a seller or a buyer,
\verb|tokens| that express the crypto-Kilowatthours (wiling to buy or to sell),
and \verb|price_order|. The \verb|order_trading| type is a record that contains
seller ID; \verb|seller_index|, buyer ID; \verb|buyer_index|, the transferred
amount \verb|amount_t|, and the trading price \verb|price_t|.

\paragraph{Remark:} In our methodology, we make the choice to encode some
primitives of \textit{Solidity} but not all. For example, the \verb|send()|
function in \textit{Solidity} can fail (return \verb|False|) due to an
out-of-gas, e.g. an overrun of 2300 units of \textit{gas}. The reason is that in
certain cases the transfer of \textit{ether} to a contract involves the
execution of the contract fallback, therefore the function might consume more
\textit{gas} than expected. A fallback function is a function without a
signature (no name, no parameters), it is executed if a contract is called and
no other function matches the specified function identifier, or if no data is
supplied. As we made the choice of a \textit{private} blockchain type, all users
can be identified and we have control on who can write or read from the
blockchain. Thus, the \textit{Why3} \verb|send()| function does not need a
fallback execution, it only transfers \textit{ether} from one address to
another. The \textit{Why3} \verb|send()| function does not return a boolean,
because we require that the transfer is possible (enough ether in the sending
contract and not too much in the receiving) and we want to avoid
Denial-of-service attack~\cite{DOS}. Indeed if we allow to propagate errors and accept to
send to untrusted contracts, it could always make our contract fail and revert.
So we can't prove any property of progress of our contract. In \textit{Tezos}
blockchain~\cite{goodman2014tezos}, call to other contracts are postponed to
after the execution of the current contract. So another contract should not be
able to make the calling contract fail.
% a transaction is built in such a way that it can not be reused.
%They also do not include fallback functions and hence they
%prevent replay attacks.

\subsubsection{Encoding and verifying functions from the BEMP application.}
\paragraph{Oracle notions.} Developping smart contracts often rely on the
concept of \textit{Oracles}~\cite{EthOracle}. An oracle can be seen as the link
between the blockchain and the ``real world". Some smart contracts functions
have arguments that are external to the blockchain. However, the blockchain does
not have access to information from an off-chain data source which is untrusted.
Accordingly, the oracle provides a service responsible for entering external
data into the blockchain, having the role of a trusted third party. However,
questions arise about the reliability of such oracles and accuracy of
information. Oracles can have unpredictable behaviour, e.g. a sensor that
measures the temperature might be an oracle, but might be faulty; thus one must
account for invalid information from oracles. 
\begin{wrapfigure}[15]{r}{0.60\textwidth}
    \centering
    \includegraphics[width=0.60\textwidth]{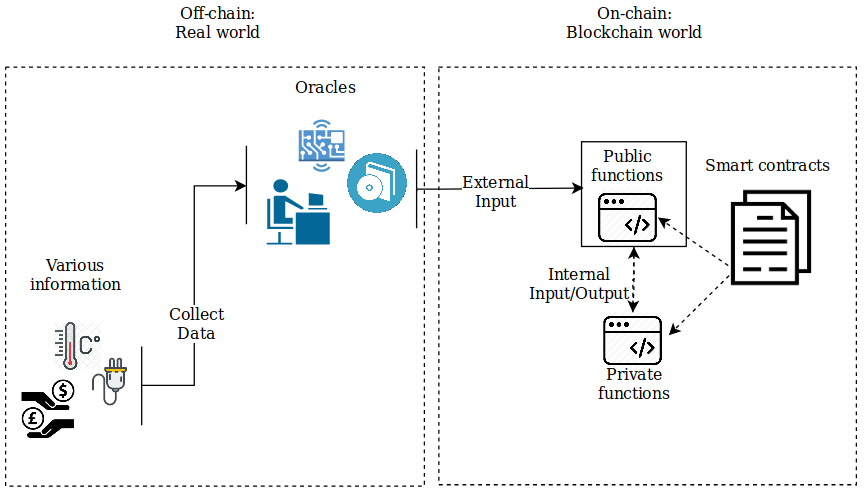}
    \caption{Link between on-chain and off-chain}
    \label{fig:modeling}
\end{wrapfigure}
Figure~\ref{fig:modeling}
illustrates the three communication stages between various systems in the real
world with the blockchain: \textit{(1)} the collection of off-chain raw data;
\textit{(2)} this data is collected by oracles; and finally, \textit{(3)}
oracles provide information to the blockchain (via smart contracts).

\paragraph{}Based on this distinction, we defined two types of functions involved in contracts, namely \textit{Private functions} and \textit{Public functions}. We noted that some functions are called  internally, by other smart contracts functions, while others are called externally by oracles. Functions that interact with oracles are defined as \textit{public} functions. The proof approach of the two types is different. For the \textit{private} functions one defines pre-conditions and post-conditions, and then we prove that no error can occur and that the function behaves as it should. It is thus not necessary to define exceptions to be raised throughout the program; they are proved to never occur. Conversely, the \textit{public} functions are called by oracles, the behaviour of the function must, therefore, take into account any input values and it is not possible to require conditions upstream of the call. So in contrast, the exceptions are necessary; we use so-called \textit{defensive proof} in order to protect ourselves from the errors that can be generated by oracles. No constraints are applied on post-conditions. Thus, valid data (which does not raise exceptions) received by a public function will satisfy the pre-conditions of the public function that uses it, because pre-conditions are proved.  
%

%\begin{figure}[t]
%    \centering
%    \includegraphics[width=0.80\textwidth]{diagramV3.png}
%    \caption{Link between on-chain and off-chain}
%    \label{fig:modeling}
%\end{figure}

\paragraph{Methodology of proving BEMP functions.}To illustrate our methodology, we take an example from BEMP. 
    \begin{lstlisting}[language=Solidity, basicstyle=\fontsize{7}{9}\tt]
function transferFromMarket(address _to, uint _value) onlyMarket returns (bool success) {
        if (exportBalanceOf[market] >= _value) 
        {/* Transferring _value from market to _to  */} 
        else {success = false;
              Error("Tokens couldn't be transferred from market");}}
\end{lstlisting}
The function allows transferring \verb|_value| (expressing cryptokwh) from the \verb|market| to \verb|_to| address. The mapping \verb|exportBalanceOf[]| stores balances corresponding to addresses that export tokens. The function can be executed solely by the market (the modifier function \verb|onlyMarket|). The program checks if the market has enough tokens to send to \verb|_to|. If this condition is verified, then the transfer is done. If the condition is not verified, the function returns \verb|false| and triggers an \verb|Error| event (a feature that allows writing logs in the blockchain)~\footnote{https://media.consensys.net/technical-introduction-to-events-and-logs-in-ethereum-a074d65dd61e}. This process is internal to the blockchain, there is no external exchange, hence the function is qualified as \textit{private}. According to the modelling approach, we define complete pre-conditions and post-conditions to verify and prove the function. The corresponding \textit{Why3} function is:   
    \begin{lstlisting}[language= Why3, basicstyle=\fontsize{7}{9}\tt]
let transferFromMarket (_to : address) (_value : uint) : bool 
    requires {!onlymarket /\ _value > 0 }
    requires {marketBalanceOf[market] >= _value }
    requires {importBalanceOf[_to] <= max_uint - _value}
    ensures {(old marketBalanceOf[market]) + (old importBalanceOf[_to]) = marketBalanceOf[market] +  importBalanceOf[_to]}
    = (* The program *)
\end{lstlisting}

The pre-condition in line 2 expresses the \verb|modifier onlyMarket| function. Note that \verb|marketBalanceOf| is the hashtable that records crypto-Kilowatthours balances associated with market addresses, and \verb|importBalanceOf| is the hashtable that records the amount of crypto-Kilowatthours intended for the buyer addresses. From the specification, we understand the behaviour of the function without referencing to the program. To be executed, \verb|transferFromMarket| must respect RTE and functional properties:
\begin{itemize}
    \item RTE properties: \textit{(1) Positive values}; a valid amount of crypto-Kilowatthours to transfer is a positive amount (Line 2). \textit{(2) Integer overflow}; no overflow will occur when \verb|_to| receives \verb|_value| (Line 4).    
    \item Functional properties:
    \textit{(1) Acceptable transfer}; the transfer can be done, if the market has enough crypto-Kilowatthours to send (Line 3).
    \textit{(2) Successful transfer}; the transaction is completed successfully if the sum of the sender and the receiver balance  before and after the execution does not change (Line 5). \textit{(3)} \verb|modifier| \textit{function}; the function can be executed only by the market (Line 2).
    
\end{itemize}
The set of specifications is necessary and sufficient to prove the expected behaviour of the function.

\paragraph{}The following function illustrates a \textit{Solidity} public function. 
    \begin{lstlisting}[language=Solidity, basicstyle=\fontsize{7}{9}\tt]
function registerSmartMeter(string _meterId, address _ownerAddress) onlyOwner {  addressOf[_meterId] = _ownerAddress;        
        MeterRegistered(_ownerAddress, _meterId);}
\end{lstlisting}
The function \verb|registerSmartMeters()| is identified by a name (\verb|meterID|) and an owner (\verb|ownerAddress|). Note that all meter owners are recorded in a hashtable \verb|addressOf| associated with a key value \verb|meterID| of the \verb|string| type. The main potential bug in this function is possibly registering a meter twice. When a meter is registered, the function broadcasts an event \verb|MeterRegistered|. Following the modelling rules, there are no pre-conditions, instead, we define exceptions. The corresponding \textit{Why3} function is:

    \begin{lstlisting}[language = Why3, basicstyle=\fontsize{7}{9}\tt]
Exception OnlyOwner, ExistingSmartMeter
let registerSmartMeter (meterID : string) (ownerAddress : address)
    raises { OnlyOwner-> !onlyOwner = False  }
    raises {ExistingSmartMeter -> mem addressOf meterID}
    ensures { (size addressOf) = (size (old addressOf) + 1 ) }
    ensures { mem addressOf meterID}
    = (*The program*)

\end{lstlisting}
The first exception (Line 3) is the \verb|modifier| function which restricts the function execution to the owner, the caller function. It is not possible to pre-condition inputs of the function, so we manage exceptional conditions during the execution of the program. To be executed, \verb|registerSmartMeter| must respect RTE and functional properties:
\begin{itemize}
    \item RTE properties: \textit{Duplicate record}; if a smart meter and its owner is recorded twice, raise an exception (Line 4) 
    \item Functional properties: \textit{(1)} \verb|modifier| \textit{function}; the function can be executed only by the owner, thus we raise \verb|OnlyOwner| when the caller of the function is not the owner (Line 3). \textit{(2) Successful record}; at the end of the function execution, we ensure (Line 5) that a record has made. \textit{(3) Existing record}; the registered smart meter has been properly recorded in the hashtable \verb|addressOf| (Line 6).
\end{itemize}
The set of specifications is necessary and sufficient to prove the expected behaviour of the function.

\paragraph{Trading contract.} The trading algorithm allows matching a potential consumer with a potential seller, recorded in two arrays \verb|buy_order| and \verb|sell_order| taken as parameters of the algorithm. In order to obtain an expected result at the end of the algorithm, properties must be respected. We define specifications that make it possible throughout the trading process. The algorithm is a private function type because it runs on-chain. Thus no exceptions are defined but pre-conditions are. The Trading contract has no \textit{Solidity} equivalent because it is a function added to the original BEMP project. Below is the set of properties of the function:
        \begin{lstlisting}[language = Why3, basicstyle=\fontsize{7}{9}\tt]
let trading (buy_order : array order) (sell_order : array order) : list order_trading
      requires { length buy_order > 0 /\ length sell_order > 0}
      requires {sorted_order buy_order}
      requires {sorted_order sell_order}
      requires {forall j:int. 0 <= j < length buy_order -> 0 < buy_order[j].tokens }
      requires {forall j:int. 0 <= j < length sell_order -> 0 < sell_order[j].tokens  }
      ensures { correct result (old buy_order) (old sell_order) }
      ensures { forall l. correct l (old buy_order) (old sell_order) ->																					 nb_token l <= nb_token result }
      ensures {!gas <= old !gas + 374 + (length buy_order + length sell_order) * 363}
      ensures {!alloc <= old !alloc + 35 + (length buy_order + length sell_order) * 35}
      = (* The program *)
\end{lstlisting}
\begin{itemize}
 \item RTE properties:\textit{ positive values}; parameters of the functions must
  not be empty (empty array) (Line 2), and a trade cannot be done with null
  or negative tokens (Lines 5, 6).\\

 \item Functional requirements: \textit{sorted orders}; the orders need to be sorted in a decreasing way. Sellers and buyers asking for the most expensive price of energy will be at the top of the list (Lines 3, 4).
  \item Functional properties: \textit{(1) correct trading} (Lines 7, 8); for a trading to be
    qualified as correct, it must satisfy two properties:
 \begin{itemize}
 \item the conservation of buyer and seller tokens that states no loss of tokens
   during the trading process : \lstinline[language = Why3,
   basicstyle=\fontsize{7}{9}\tt]
   ! forall i:uint. 0 <= i < length sell_order -> sum_seller (list_trading) i <= sell_order[i].tokens!.
   For the buyer it is
   equivalent by replacing seller by buyer.
 \item a successful matching; a match between a seller and a buyer is qualified
   as correct if the price offered by the seller is less than or equal to that
   of the buyer, and that the sellers and buyers are valid indices in the array.
 \end{itemize}
 \textit{(2) Best tokens exchange}; we choose to qualify a trade as
 being one of the best if it maximize the total number of tokens exchanged.
 Line 8 ensures that no correct trading list can have more tokens exchanged than the one resulting from the function. The criteria could be refined by adding that we then want to maximize or minimize the sum of paid (best for seller or for buyer). \textit{(3) Gas consumption}; Lines 9 and 10 ensures that no extra-consumption of gas will happen (see the following paragraph).
\end{itemize}
\paragraph{Gas consumption proof.} Overconsumption of \textit{gas} can be avoided by the \textit{gas} model. Instructions in EVM consume an amount of \textit{gas}, and they are categorized by level of difficulty; e.g., for the set $W_{verylow}=\{ ADD,\ SUB,\ ...\}$, the amount to pay is $G_{verylow} =\ 3\ units\ of\ gas$, and for a create operation the amount to pay is $G_{create} =\ 32000\ units\ of\ gas$~\cite{wood2014ethereum}. The price of an operation is proportional to its difficulty. Accordingly, we fix for each \textit{Why3} function, the appropriate amount of \textit{gas} needed to execute it. Thus, at the end of the function instructions, a variable \verb|gas| expresses the total quantity of \textit{gas} consumed during the process. We introduce a \verb|val ghost| function that adds to the variable \verb|gas| the amount of \textit{gas} consumed by each function calling \verb|add_gas| (see section 3 for more details on \textit{gas} allocation).
\begin{lstlisting}[language = Why3, basicstyle=\fontsize{7}{9}\tt]
val ghost add_gas (used : gas) (allocation: int): unit
    requires { 0 <= used /\ 0 <= allocation }
    ensures  { !gas = (old !gas) + used }
    ensures  { !alloc = (old !alloc) + allocation }
    writes   { gas, alloc}
    
\end{lstlisting}

\paragraph{}The specifications of the function above require \textit{positive values} (Line 2). Moreover, at the end of the function, we ensure that there is no extra \textit{gas} consumption (Lines 3, 4). Line 5 specifies the changing variables.

%Moreover, we introduce a \verb|max_gas| value which defines the maximum amount that can be consumed at most during the whole process of a given program. Accordingly, \verb|max_gas| is necessarily positive. Furthermore, a predicate allow verifying that the program has consumed at most \verb|max_gas| : \lstinline[language = Why3, basicstyle=\fontsize{7}{9}\tt] ! predicate consumption_at_most(amount : gas) = max_gas - amount >= 0!.

\section{Compiling Why3 Contracts and Proving Gas Consumption}
The final step of the approach is the
deployment of \textit{Why3} contracts. EVM is designed to be the runtime
environment for the smart contracts on the Ethereum
blockchain~\cite{wood2014ethereum}.
% Smart contracts are just like
% regular accounts, except they run EVM bytecode when receiving a
% transaction, allowing them to perform calculations and further
% transactions.
The EVM is a stack-based machine (word of 256 bits) and uses a set of
instructions (called opcodes)\footnote{https://ethervm.io} to execute specific tasks. The EVM features two memories, one volatile that does not survive the current transaction and a second for storage that does survive but is a lot more expensive to modify. The goal of this section is to describe the approach of compiling \textit{Why3} contracts into EVM code and proving the cost of functions. The compilation\footnote{The implementation can be found at
  \url{http://francois.bobot.eu/fm2019/}} is done in three phases: \textit{(1)}
compiling to an EVM that uses symbolic labels for jump destination and macro
instructions. \textit{(2)} computing the absolute address of the labels, it must
be done inside a fixpoint because the size of the jump addresses has an impact
on the size of the instruction. Finally, \textit{(3)} translating the assembly
code to pure EVM assembly and printed. Most of \textit{Why3} can be translated,
the proof-of-concept compiler allows using algebraic datatypes, not nested
pattern-matching, mutable records, recursive functions, while loops, integer
bounded arithmetic (32, 64,128, 256 bits). Global variables are restricted to
mutable records with fields of integers. It could be extended to hashtables
using the hashing technique of the keys used in \textit{Solidity}. Without using
specific instructions, like for C, \textit{Why3} is extracted to garbage
collected language, here all the allocations are done in the volatile memory, so
the memory is reclaimed only at the end of the transaction.

\paragraph{} We have not formally proved yet the correction of the compilation, we only tested the compiler using reference interpreter \cite{} and by asserting some invariants during the transformation.
However, we could list the following arguments for the correction:
\begin{itemize}
\item the compilation of why3 (ML-language) is straightforward to stack machine.
  % The only optimization is that Why3 unit expressions are translated to EVM code
  % that 
\item the precondition on all the arithmetic operations (always bounded) ensures
  arithmetic operations could directly use 256bit operations
\item raise accepted only in public function before any
  mutation so the fact they are translated into revert does not change their semantics. \lstinline{try with} are forbidden.
\item only immutable datatype can be stored in the permanent store. Currently, only integers can be stored, it could be extended to other immutable datatye
  by copying the data to and from the store.
\item The send function in why3 only modifies the state of balance of the
  contracts, requires that the transfer is acceptable and never fail, as
  discussed previously. So it is
  compiled similarly to the solidity function send function with a gas limit
  small enough to disallow modification of the store. Additionally, we discard the
  result.
\end{itemize}

The execution of
each bytecode instruction has an associated cost. One must pay some \textit{gas}
when sending a transaction; if there is not enough \textit{gas} to execute the
transaction, the execution stops and the state is rolled back. So it is
important to be sure that at any later date the execution of a smart contract
will not require an unreasonable quantity of \textit{gas}. The computation of
WCET is facilitated in EVM by the absence of cache. So we could use techniques
of~\cite{cerco} which annotate in the source code the quantity of \textit{gas}
used, here using a function \lstinline{add_gas used allocations}. The number of
allocations is important because the real \textit{gas} consumption of EVM
integrates the maximum quantity of volatile memory used. The compilation checks
that all the paths of the function have a cost smaller than the sum of the
\lstinline{add_gas g a} on it. The paths of a function are defined on the EVM
code by starting at the function-entry and loop-head and going through the code
following jumps that are not going back to loop-head.
\begin{center}
\begin{minipage}{0.95\textwidth}

% \begin{lstlisting}[language = Why3, basicstyle=\fontsize{7}{9}\tt]
% let rec length4 [@ evm:gas_checking] (l:list 'a) : int32
%  requires { (length l) <= max_int32 }
%  ensures { !gas - old !gas <= (length l) * 128 + 71 }
%  ensures { !alloc - old !alloc <= 0 }
%  ensures { result = length l }
%  variant { l } =
%   match l with
%   | Nil -> add_gas 71 0; 0
%   | Cons _ l -> add_gas 128 0; 1 + length4 l
%   end

% let rec mk_list42 [@ evm:gas_checking] (i:int32) : list int32
%   requires { 0 <= i }
%   ensures { !gas - old !gas <= i * 185 + 113 }
%   ensures { !alloc - old !alloc <= i * 96 + 32 }
%   ensures { i = length result }
%   variant { i } =
%   if i <= 0 then (add_gas 113 32; Nil) else
%   let l = mk4 (i-1) in
%   add_gas 185 96;
%   Cons (0x42:int32) l

% let g4 [@ evm:gas_checking] (i:int32) : int32
%   requires { 0 <= i }
%   ensures { !gas - old !gas <= i * 313 + 242 }
%   ensures { !alloc - old !alloc <= i * 96 + 32 }
% =
% add_gas 58 0;
% let l = mk4 i in
% length4 l
% \end{lstlisting}
\begin{lstlisting}[language = Why3, basicstyle=\fontsize{7}{9}\tt]
let rec mk_list42 [@ evm:gas_checking] (i:int32) : list int32
 requires { 0 <= i }  ensures { i = length result }  variant { i }
 ensures { !gas - old !gas <= i * 185 + 113 }
 ensures { !alloc - old !alloc <= i * 96 + 32 } =
   if i <= 0 then (add_gas 113 32; Nil)
   else (let l = mk_list42 (i-1) in add_gas 185 96; Cons (0x42:int32) l)
\end{lstlisting}
\end{minipage}
\end{center}
% Then we can lift this information using Why3
% specification to prove that a function that given \lstinline{i} builds
% a list of length \lstinline{l} has a cost smaller than $185 i + 113$
% and allocates at most $96 i + 32$ bytes.
\paragraph{}Currently, the cost of the modification of storage is
over-approximated; using specific contract for the functions that modify it we
could specify that it is less expansive to use a memory cell already used.

\section{Related Work}
Since the \textit{DAO} attack, the introduction of formal methods at the level of smart contracts has increased. Raziel is a framework to prove the validity of smart contracts to third parties before their execution in a private way~\cite{sanchez2017raziel}. In that paper, the authors also use a deductive proof approach, but their concept is based on Proof-Carrying Code (PCC) infrastructure, which consists of annotating the source code, thus proofs can be checked before contract execution to verify their validity. Our method does not consist in annotating the \textit{Solidity} source code but in writing the contract program and thus getting a correct-by-construction program. Another widespread approach is static analysis tools. One of them is called Oyente. It has been developed to analyze Ethereum smart contracts to detect bugs. In the corresponding paper~\cite{luu2016making}, the authors were able to run Oyente on 19,366 existing Ethereum contracts, and as a result, the tool flagged 8,833 of them as vulnerable. Although that work provides interesting conclusions, it uses symbolic execution, analyzing paths, so it does not allow to prove functional properties of the entire application. We can also mention the work undertaken by the \textit{F*} community~\cite{bhargavan2016short} where they use their functional programming language to translate \textit{Solidity} contracts to shallow-embedded F* programs. Just like~\cite{ahrendtverification} where the authors perform static analysis by translating \textit{Solidity} contracts into Java using \textit{KeY}~\cite{ahrendt2016deductive}. The initiative of the current paper is directly related to a previous work~\cite{nehai}, which dealt with formally verifying the smart contracts application by using model-checking. The paper established a methodology to construct a three-fold model of an Ethereum application, with properties formalized in temporal logic CTL. However, because of the limitation of the model-checker used, ambitious verification could not be achieved (e.g., a model for $m$ consumers and $n$ producers). This present work aims to surpass the limits encountered with  model-checking, by using a deductive proof approach on an Ethereum application using the \textit{Why3} tool.

\section{Conclusions}
In this paper, we applied concepts of deductive verification to a computer protocol intended  to enforce some transaction rules within an Ethereum blockchain application. The aim is to avoid errors that could have serious consequences. Reproducing, with \textit{Why3}, the behaviour of \textit{Solidity} functions showed that \textit{Why3} is suitable for writing and verifying smart contracts programs. 
%In this theorem proving approach, we define mathematical statements to be proved as pre-conditions, post-conditions, invariants, etc.
%Furthermore, because the Solidity language contains elements which are not part of the Why3 language, we built a Why3 library dedicated to Solidity expressions, thus the encoding from one language to another is more trustworthy and accurate. 
The presented method was applied to a use case that describes an energy market place allowing local energy trading among inhabitants of a neighbourhood. The resulting modelling allows establishing a trading contract, in order to match consumers with producers willing to make a transaction. In addition, this last point demonstrates that with a deductive approach it is possible to model and prove the operation of the BEMP application at realistic scale (e.g. matching $m$ consumers with $n$ producers), contrary to model-checking in~\cite{nehai}, thus allowing the verifying of more realistic functional properties.
%\paragraph{}Future work will focus on verifying and proving blockchain properties involving communication between users and the impact that a lack of \textit{ether} can have on the entire eco-system of the blockchain.  
\newpage

% The reliability of our translation process from Solidity program to a Why3 program comes from the fact that the equivalence between the two languages is not concerned because the goal is to write a program in why3, sufficiently sure and check it, so that we can then send the proved code directly to an EVM compiler, so we do not try to have a proper translation. We want to break away from the Solidity language because it is a source of bug. The difficulties encountered in translating to why3 is to find the right invariant for each loop of the program, because they must be precise and write manually. Another problem encountered with the Why3 tool is the mutually recursive modules, therefore in some cases, it is necessary to divide the modules that model a contract because a function call cycle occur, and the tool does not allow that. it allows it with the functions if recursively is announced. 
\newpage

\bibliographystyle{splncs04}
\bibliography{biblio}

\begin{thebibliography}{10}
\providecommand{\url}[1]{\texttt{#1}}
\providecommand{\urlprefix}{URL }
\providecommand{\doi}[1]{https://doi.org/#1}

\bibitem{EthOracle}
Ethereum foundation : Ethereum and oracles.
  \url{https://blog.ethereum.org/2014/07/22/ethereum-and-oracles/}

\bibitem{SolandWhy}
Formal verification for solidity contracts.
  \url{https://forum.ethereum.org/discussion/
  3779/formal-verification-for-solidity-contracts}

\bibitem{DOS}
Solidity hacks and vulnerabilities.
  \url{https://hackernoon.com/hackpedia-16-solidity-hacks-vulnerabilities-their-fixes-and-real-world-examples-f3210eba5148}

\bibitem{ahrendt2016deductive}
Ahrendt, W., Beckert, B., Bubel, R., H{\"a}hnle, R., Schmitt, P.H., Ulbrich,
  M.: Deductive software verification-the key book. lncs, vol. 10001 (2016)

\bibitem{ahrendtverification}
Ahrendt, W., Bubel, R., Ellul, J., Pace, G.J., Pardo, R., Rebiscoul, V.,
  Schneider, G.: Verification of smart contract business logic  (2019)

\bibitem{cerco}
Amadio, R.M., Ayache, N., Bobot, F., Boender, J.P., Campbell, B., Garnier, I.,
  Madet, A., McKinna, J., Mulligan, D.P., Piccolo, M., Pollack, R.,
  R{\'e}gis-Gianas, Y., Sacerdoti~Coen, C., Stark, I., Tranquilli, P.:
  Certified complexity (cerco). In: Dal~Lago, U., Pe{\~{n}}a, R. (eds.)
  Foundational and Practical Aspects of Resource Analysis. pp. 1--18. Springer
  International Publishing, Cham (2014)

\bibitem{atzei2017survey}
Atzei, N., Bartoletti, M., Cimoli, T.: A survey of attacks on ethereum smart
  contracts. In: Principles of Security and Trust, pp. 164--186. Springer
  (2017)

\bibitem{BCD+11}
Barrett, C., Conway, C.L., Deters, M., Hadarean, L., Jovanovi\'{c}, D., King,
  T., Reynolds, A., Tinelli, C.: {CVC4}. In: Proceedings of the 23rd
  International Conference on Computer Aided Verification. Springer (2011)

\bibitem{bhargavan2016short}
Bhargavan, K., Delignat-Lavaud, A., Fournet, C., Gollamudi, A., Gonthier, G.,
  Kobeissi, N., Rastogi, A., Sibut-Pinote, T., Swamy, N., Zanella-B{\'e}guelin,
  S.: Short paper: Formal verification of smart contracts (2016)

\bibitem{altergo}
Bobot, F., Conchon, S., Contejean, E., Iguernelala, M., Lescuyer, S., Mebsout,
  A.: The alt-ergo automated theorem prover (2008),
  \url{http://alt-ergo.lri.fr/}

\bibitem{buterin2014next}
Buterin, V., et~al.: A next-generation smart contract and decentralized
  application platform. white paper  (2014)

\bibitem{domowitz1993taxonomy}
Domowitz, I.: A taxonomy of automated trade execution systems. Journal of
  International Money and Finance  \textbf{12},  607--631 (1993)

\bibitem{filliatre2013why3}
Filli{\^a}tre, J.C., Paskevich, A.: Why3 -- where programs meet provers. In:
  European Symposium on Programming. pp. 125--128. Springer (2013)

\bibitem{goodman2014tezos}
Goodman, L.: Tezos: A self-amending crypto-ledger position paper (2014)

\bibitem{luu2016making}
Luu, L., Chu, D.H., Olickel, H., Saxena, P., Hobor, A.: Making smart contracts
  smarter. In: Proceedings of the 2016 ACM SIGSAC Conference on Computer and
  Communications Security. pp. 254--269. ACM (2016)

\bibitem{z3smt}
de~Moura, L., Bj{\o}rner, N.: Z3, an efficient {SMT} solver,
  \url{http://research.microsoft.com/projects/z3/}

\bibitem{nakamoto2008bitcoin}
Nakamoto, S.: Bitcoin: A peer-to-peer electronic cash system  (2008)

\bibitem{nehai}
Neha\"{\i}, Z., Piriou, P.Y., Daumas, F.: Model-checking of smart contracts.
  In: The 2018 IEEE International Conference on Blockchain. IEEE (2018)

\bibitem{sanchez2017raziel}
S{\' a}nchez, D.C.: Raziel: Private and verifiable smart contracts on
  blockchains. Cryptology ePrint Archive, Report 2017/878 (2017),
  \url{http://eprint.iacr.org/2017/878.pdf}, accessed:2017-09-26

\bibitem{wood2014ethereum}
Wood, G.: Ethereum: A secure decentralised generalised transaction ledger.
  Ethereum project yellow paper  \textbf{151},  1--32 (2014)

\end{thebibliography}
\newpage
\begin{subappendices}
\renewcommand{\thesection}{\Alph{section}}
\section{: BEMP Application}
\label{appendix:algo}

  \centering
\begin{lstlisting}[language=why3, basicstyle=\small\tt ]
module DCC (*the module that materializes the smart meters*)
  use my_library.Uint
  use my_library.SmartMeterID
  use my_library.Address
  use array.Array

    (*records of potential selleur and buyeur, with the purchase (price_b) and sale (price_s) price*)
    (*amount_b the needed token quantity, and amount_s the token quantity on sale*)

    type pot_buy = {address_b : address;
                    smb_id: smartMeterID;
                    price_b: uint;
                    amount_b: uint}

    type pot_sell = {address_s : address;
                     sms_id : smartMeterID;
                     price_s: uint;
                     amount_s: uint}


    (*buy_array and sell_array are data tables retrieved from the meters*)
    val buy_array : array pot_buy
    val sell_array : array pot_sell


end

module Trading
  use my_library.Uint
  use int.Int
  use int.MinMax
  use seq.Seq
  use import my_library.ArrayUint as Arr
  use ref.Refint
  use list.List
  use import list.Length as Len
  use list.NthNoOpt
  use my_library.SmartMeterID
  use my_library.Address
  use list.HdTlNoOpt
  use list.NthHdTl
  use list.Nth as Elem


    type order = {orderAddress : address; tokens: uint; price_order: uint} (*It can be buy or sell , tokens = energy materializes in token*)

    clone array.Sorted as Sort with type elt = order

    val sorted_array (a: array order) : unit
      ensures {forall i j: int. 0 <= j <= i < Arr.length a -> Uint.to_int(a[i].price_order) <= Uint.to_int(a[j].price_order)}
      writes {a}


    predicate sorted_order (a: Seq.seq order) =
      forall k1 k2 : int. 0 <= k1 <= k2 < Seq.length a ->
      Uint.to_int(a[k2].price_order) <= Uint.to_int(a[k1].price_order)

(**)

    type order_trading = {seller_index: uint; buyer_index: uint; amount_t: uint}

    predicate matching_order (k: order_trading) (b_order : Seq.seq order) (s_order : Seq.seq order) =
              s_order[k.seller_index].price_order <=
              b_order[k.buyer_index].price_order  /\
              0 <= k.buyer_index < Seq.length b_order /\
              0 <= k.seller_index < Seq.length s_order /\
              0 < k.amount_t

    predicate matching (order: list order_trading) (b_order : Seq.seq order) (s_order : Seq.seq order) =
       match order with
       | Nil -> true
       | Cons k l -> matching l b_order s_order /\
                     matching_order k b_order s_order
       end

   let rec lemma matching_nth (order: list order_trading) (b_order : Seq.seq order) (s_order : Seq.seq order)
      requires { matching order b_order s_order }
      ensures { forall k :int. 0 <= k < Len.length order ->
                 matching_order (nth k order) b_order s_order }
      variant { order }
                  =
      match order with
      | Nil -> ()
      | Cons _ l -> matching_nth l b_order s_order
      end                 

   let rec lemma matching_same_price (order: list order_trading) (b_order : Seq.seq order) (s_order : Seq.seq order) (b_order' : Seq.seq order) (s_order' : Seq.seq order)
      requires { matching order b_order s_order }
      requires { Seq.length b_order = Seq.length b_order' }
      requires { Seq.length s_order = Seq.length s_order' }
      requires {forall j:int. 0 <= j < Seq.length b_order -> b_order'[j].price_order = b_order[j].price_order }
      requires {forall j:int. 0 <= j < Seq.length s_order -> s_order'[j].price_order = s_order[j].price_order }
      ensures { matching order b_order' s_order' }
      variant { order }
                  =
      match order with
      | Nil -> ()
      | Cons _ l ->
        matching_same_price l b_order s_order b_order' s_order'
      end                 


    predicate smallest_buyer_seller (order: list order_trading) (buyer : int) (seller : int) =
       match order with
       | Nil -> true
       | Cons k l -> smallest_buyer_seller l buyer seller /\
                     k.buyer_index >= buyer /\
                     k.seller_index >= seller
       end


    function sum_seller (l : list order_trading) (sellerIndexe : int) : int
    =
    match l with
    | Nil -> 0
    | Cons h t -> ( if  h.seller_index = sellerIndexe then Uint.to_int(h.amount_t) else 0 ) + sum_seller t sellerIndexe
    end

    let rec lemma sum_seller_positive (l : list order_trading) (buyerIndexe : int)
     ensures { 0 <= sum_seller l buyerIndexe }
     =
      match l with
      | Nil -> ()
      | Cons _ l -> sum_seller_positive (l : list order_trading) (buyerIndexe : int)
      end

    function sum_buyer (l : list order_trading) (buyerIndexe : int) : int
    =
    match l with
    | Nil -> 0
    | Cons h t -> ( if  h.buyer_index = buyerIndexe then Uint.to_int(h.amount_t) else 0 ) + sum_buyer t buyerIndexe     end


    let rec lemma sum_buyer_positive (l : list order_trading) (buyerIndexe : int)
      ensures { 0 <= sum_buyer l buyerIndexe }
     =
      match l with
      | Nil -> ()
      | Cons _ l -> sum_buyer_positive (l : list order_trading) (buyerIndexe : int)
      end
      
    let rec lemma smallest_buyer_seller_sum_seller (order: list order_trading) (buyer : int) (seller : int)  (b_order : Seq.seq order) (s_order : Seq.seq order)
        requires { matching order b_order s_order }
        requires { smallest_buyer_seller order buyer seller }
        requires { sum_seller order seller = 0 }
        ensures { smallest_buyer_seller order buyer (seller + 1) }
      =
       match order with
       | Nil -> ()
       | Cons _ l ->
          smallest_buyer_seller_sum_seller (l: list order_trading) (buyer : int) (seller : int) b_order s_order
       end

    let rec lemma smallest_buyer_seller_sum_buyer (order: list order_trading) (buyer : int) (seller : int)  (b_order : Seq.seq order) (s_order : Seq.seq order)
        requires { matching order b_order s_order }
        requires { smallest_buyer_seller order buyer seller }
        requires { sum_buyer order buyer = 0 }
        ensures { smallest_buyer_seller order (buyer + 1) seller }
      =
       match order with
       | Nil -> ()
       | Cons _ l ->
          smallest_buyer_seller_sum_buyer (l: list order_trading) (buyer : int) (seller : int) b_order s_order
       end

    let rec lemma smallest_buyer_seller_expensive_seller (order: list order_trading) (buyer : int) (seller : int)  (b_order : Seq.seq order) (s_order : Seq.seq order)
        requires { matching order b_order s_order }
        requires { sorted_order b_order }
        requires { 0 <= buyer < Seq.length b_order }
        requires { smallest_buyer_seller order buyer seller }
        requires { b_order[buyer].price_order < s_order[seller].price_order }
        ensures { smallest_buyer_seller order buyer (seller + 1) }
        variant { order }
      =
       match order with
       | Nil -> ()
       | Cons _ l ->
            smallest_buyer_seller_expensive_seller (l: list order_trading) (buyer : int) (seller : int) b_order s_order
       end

    let lemma smallest_buyer_seller_after_last (order: list order_trading) (buyer : int) (seller : int)  (b_order : Seq.seq order) (s_order : Seq.seq order)
        requires { matching order b_order s_order }
        requires { smallest_buyer_seller order buyer seller }
        requires { Seq.length s_order <= seller \/ Seq.length b_order <= buyer }
        ensures { order = Nil }
      =
       match order with
       | Nil -> ()
       | Cons _ _ ->
         absurd
       end


    function nb_token (l : list order_trading) : int
    =
    match l with
    | Nil -> 0
    | Cons h t -> h.amount_t + nb_token t
    end

    let rec lemma nb_token_positive (l : list order_trading)
      ensures { 0 <= nb_token l}
     =
      match l with
      | Nil -> ()
      | Cons _ l -> nb_token_positive (l : list order_trading)
      end

    let rec lemma nb_token_zero_sum_buyer (l : list order_trading) (indexe : uint)
      requires { nb_token l = 0 }
      ensures { sum_seller l indexe = 0 }
      ensures { sum_buyer  l indexe = 0 }
     =
      match l with
      | Nil -> ()
      | Cons _ l -> nb_token_zero_sum_buyer (l : list order_trading) (indexe : uint)
      end

    predicate correct (l:list order_trading) (buy_order: Seq.seq order) (sell_order: Seq.seq order) =
      (forall i:uint. 0 <= i < Seq.length sell_order ->
                  sum_seller l i <=  Uint.to_int(sell_order[i].tokens)) /\
      (forall i:uint. 0 <= i < Seq.length buy_order ->
                  sum_buyer l i <=  Uint.to_int(buy_order[i].tokens)) /\
       matching l buy_order sell_order

    let rec ghost find_seller (l:list order_trading) (buy_order: Seq.seq order) (sell_order: Seq.seq order) (buyer:uint) (seller:uint) : (list order_trading , order_trading)
        requires { matching l buy_order sell_order }
        requires { smallest_buyer_seller l buyer seller }
        requires { 0 < sum_seller l seller }
        ensures { let l',_ = result in nb_token l = 1 + nb_token l' }
        ensures { let l',k = result in
                    forall buyer. sum_buyer l buyer = sum_buyer l' buyer + (if k.buyer_index = buyer then 1 else 0)  }
        ensures { let l',k = result in
                    forall seller. sum_seller l seller = sum_seller l' seller + (if k.seller_index = seller then 1 else 0)  }
        ensures { let l',_ = result in matching l' buy_order sell_order }
        ensures { let l',_ = result in smallest_buyer_seller l' buyer seller }
        ensures { let _,k = result in k.seller_index = seller }
        ensures { let _,k = result in k.buyer_index >= buyer }
        ensures { let _,k = result in matching_order k buy_order sell_order }
        variant { l }
    =
      match l with
      | Nil -> absurd
      | Cons k l ->
       if k.seller_index = seller then
        if k.amount_t = 1 then l,k else (Cons {k with amount_t = k.amount_t - 1} l), {k with amount_t = 1} 
       else
        let l,k' = find_seller l buy_order sell_order buyer seller in
        (Cons k l),k'
      end

    let rec ghost find_buyer (l:list order_trading) (buy_order: Seq.seq order) (sell_order: Seq.seq order) (buyer:uint) (seller:uint) : (list order_trading , order_trading)
        requires { matching l buy_order sell_order }
        requires { smallest_buyer_seller l buyer seller }
        requires { 0 < sum_buyer l buyer }
        ensures { let l',_ = result in nb_token l = 1 + nb_token l' }
        ensures { let l',k = result in
                    forall buyer. sum_buyer l buyer = sum_buyer l' buyer + (if k.buyer_index = buyer then 1 else 0)  }
        ensures { let l',k = result in
                    forall seller. sum_seller l seller = sum_seller l' seller + (if k.seller_index = seller then 1 else 0)  }
        ensures { let l',_ = result in matching l' buy_order sell_order }
        ensures { let l',_ = result in smallest_buyer_seller l' buyer seller }
        ensures { let _,k = result in k.buyer_index = buyer }
        ensures { let _,k = result in k.seller_index >= seller }
        ensures { let _,k = result in matching_order k buy_order sell_order }
        variant { l }
    =
      match l with
      | Nil -> absurd
      | Cons k l ->
       if k.buyer_index = buyer then
        if k.amount_t = 1 then l,k else (Cons {k with amount_t = k.amount_t - 1} l), {k with amount_t = 1} 
       else
        let l,k' = find_buyer l buy_order sell_order buyer seller in
        (Cons k l),k'
      end

    let ghost remove_seller_buyer_token1 (l:list order_trading) (buy_order: Seq.seq order) (sell_order: Seq.seq order) (buyer:uint) (seller:uint) : list order_trading
        requires { sorted_order buy_order }
        requires { sorted_order sell_order }
        requires { matching l buy_order sell_order }
        requires { smallest_buyer_seller l buyer seller }
        requires { 1 <= sum_seller l seller }
        requires { 1 <= sum_buyer l buyer }
        requires { buy_order[buyer].price_order >= sell_order[seller].price_order }
        ensures { nb_token l = 1 + nb_token result }
        ensures { forall buyer'. sum_buyer l buyer' = sum_buyer result buyer' + (if buyer' = buyer then 1 else 0)  }
        ensures { forall seller'. sum_seller l seller' = sum_seller result seller' + (if seller' = seller then 1 else 0)  }
        ensures { matching result buy_order sell_order }
        ensures { smallest_buyer_seller result buyer seller }
    =
      let l, k = find_seller l buy_order sell_order buyer seller in
      if k.buyer_index = buyer then l
      else
       let l, k' = find_buyer l buy_order sell_order buyer seller in
       assert { buy_order[k.buyer_index].price_order >= sell_order[seller].price_order };
       assert { buy_order[buyer].price_order >= sell_order[k'.seller_index].price_order };
       Cons { buyer_index = k.buyer_index; seller_index = k'.seller_index; amount_t = 1 } l

   let ghost remove_seller_token1 (l:list order_trading) (buy_order: Seq.seq order) (sell_order: Seq.seq order) (buyer:uint) (seller:uint) : list order_trading
        requires { sorted_order buy_order }
        requires { sorted_order sell_order }
        requires { matching l buy_order sell_order }
        requires { smallest_buyer_seller l buyer seller }
        requires { 1 <= sum_seller l seller }
        requires { buy_order[buyer].price_order >= sell_order[seller].price_order }
        ensures { nb_token l = 1 + nb_token result }
        ensures { forall buyer'. sum_buyer l buyer' >= sum_buyer result buyer' }
        ensures { forall seller'. sum_seller l seller' = sum_seller result seller' + (if seller' = seller then 1 else 0)  }
        ensures { matching result buy_order sell_order }
        ensures { smallest_buyer_seller result buyer seller }
    =
      let l,_ = find_seller l buy_order sell_order buyer seller in
      l
      
    let ghost remove_buyer_token1 (l:list order_trading) (buy_order: Seq.seq order) (sell_order: Seq.seq order) (buyer:uint) (seller:uint) : list order_trading
        requires { sorted_order buy_order }
        requires { sorted_order sell_order }
        requires { matching l buy_order sell_order }
        requires { smallest_buyer_seller l buyer seller }
        requires { 1 <= sum_buyer l buyer }
        requires { buy_order[buyer].price_order >= sell_order[seller].price_order }
        ensures { nb_token l = 1 + nb_token result }
        ensures { forall buyer'. sum_buyer l buyer' = sum_buyer result buyer' + (if buyer' = buyer then 1 else 0) }
        ensures { forall seller'. sum_seller l seller' >= sum_seller result seller' }
        ensures { matching result buy_order sell_order }
        ensures { smallest_buyer_seller result buyer seller }
    =
      let l,_ = find_buyer l buy_order sell_order buyer seller in
      l


   let rec ghost remove_token1 (l:list order_trading) (buy_order: Seq.seq order) (sell_order: Seq.seq order) (buyer:uint) (seller:uint) : list order_trading
        requires { sorted_order buy_order }
        requires { sorted_order sell_order }
        requires { matching l buy_order sell_order }
        requires { smallest_buyer_seller l buyer seller }
        requires { buy_order[buyer].price_order >= sell_order[seller].price_order }
        requires { 0 < nb_token l }
        ensures { nb_token l = 1 + nb_token result }
        ensures { forall buyer'. sum_buyer l buyer' >= sum_buyer result buyer' }
        ensures { forall seller'. sum_seller l seller' >= sum_seller result seller' }
        ensures { matching result buy_order sell_order }
        ensures { smallest_buyer_seller result buyer seller }
        variant { l }
    =
      match l with
      | Nil -> absurd
      | Cons k l ->
      if k.amount_t = 1 then l
      else Cons { k with amount_t = k.amount_t - 1 } l
      end


    let rec ghost remove_seller_buyer' (l:list order_trading) (buy_order: Seq.seq order) (sell_order: Seq.seq order) (buyer:uint) (seller:uint) (token: uint) : list order_trading
        requires { sorted_order buy_order }
        requires { sorted_order sell_order }
        requires { matching l buy_order sell_order }
        requires { smallest_buyer_seller l buyer seller }
        requires { buy_order[buyer].price_order >= sell_order[seller].price_order }
        ensures { nb_token l <= token + nb_token result }
        ensures { forall buyer'. buyer' <> buyer -> sum_buyer l buyer' >= sum_buyer result buyer' }
        ensures { forall seller'. seller' <> seller -> sum_seller l seller' >= sum_seller result seller' }
        ensures { max (sum_buyer l buyer - token) 0 = sum_buyer result buyer }
        ensures { max (sum_seller l seller - token) 0 = sum_seller result seller }
        ensures { matching result buy_order sell_order }
        ensures { smallest_buyer_seller result buyer seller }
        variant { token }
        writes { }
        reads { }
    =
      if token = 0 then l
      else
        let l =
           if 0 < sum_seller l (Uint.to_int seller) && 0 < sum_buyer l (Uint.to_int buyer)
           then remove_seller_buyer_token1 l buy_order sell_order buyer seller
           else if 0 < sum_seller l (Uint.to_int seller) then
           remove_seller_token1 l buy_order sell_order buyer seller
           else if 0 < sum_buyer l (Uint.to_int buyer) then
           remove_buyer_token1 l buy_order sell_order buyer seller
           else if 0 < nb_token l then
           remove_token1 l buy_order sell_order buyer seller
           else
           Nil
        in
        remove_seller_buyer' l buy_order sell_order buyer seller (token-1)

    (* Trading algorithm that matches sales and purchases *)
    (* as input I have an array of buy orders and an array of sell orders *)
    let trading (buy_order : array order) (sell_order : array order) : list order_trading
      requires { Arr.length buy_order > 0 /\ Arr.length sell_order > 0}
      requires {sorted_order buy_order}
      requires {sorted_order sell_order}
      requires {forall j:int. 0 <= j < Arr.length buy_order -> 0 < buy_order[j].tokens }
      requires {forall j:int. 0 <= j < Arr.length sell_order -> 0 < sell_order[j].tokens  }
      ensures { correct result (old buy_order) (old sell_order) }
      ensures { forall l. correct l (old buy_order) (old sell_order) ->
                                  nb_token l <= nb_token result }
      =
        (*order_list the output of the function*)
        (*order list that brings together the matching between seller and buyer*)
        let order_list : ref (list order_trading) = ref Nil in
        let i = ref (0:uint) in
        let j = ref (0:uint) in

          (*I sort my arrays in a decreasing way*)
          assert{sorted_order buy_order};
          label Before in

        let ghost others = ref (fun (l:list order_trading) -> l) in
        let ghost buy_order0 = pure { buy_order.elts } in
        let ghost sell_order0 = pure { sell_order.elts } in

        while Uint.(<) !i (Arr.length buy_order) && Uint.(<) !j (Arr.length sell_order) do

         invariant {0 <= !i <= Arr.length (buy_order at Before) /\ 0 <= !j <= Arr.length (sell_order at Before)}
         invariant {0 <= !i <= Arr.length (buy_order) /\ 0 <= !j <= Arr.length (sell_order )}
         invariant {sorted_order (buy_order at Before)}
         invariant {sorted_order (sell_order at Before)}

         invariant {forall j: int. 0 <= j < Arr.length buy_order -> buy_order[j].orderAddress == (buy_order[j].orderAddress at Before)}
         invariant {forall j: int. 0 <= j < Arr.length sell_order -> sell_order[j].orderAddress == (sell_order[j].orderAddress at Before)}

         invariant {forall j:int. 0 <= j < Arr.length (buy_order at Before) -> (buy_order at Before)[j].price_order = buy_order[j].price_order }
         invariant {forall j:int. 0 <= j < Arr.length (sell_order at Before) -> (sell_order at Before)[j].price_order = sell_order[j].price_order }

         invariant {forall j:int. 0 <= j < Arr.length (buy_order at Before) -> Uint.to_int(buy_order[j].tokens) <= Uint.to_int((buy_order at Before)[j].tokens) }
         invariant {forall j:int. 0 <= j < Arr.length (sell_order at Before) -> Uint.to_int(sell_order[j].tokens) <= Uint.to_int((sell_order at Before)[j].tokens)  }

         invariant {forall k:int. !i <= k < Arr.length (buy_order at Before) -> 0 < Uint.to_int(buy_order[k].tokens) }
         invariant {forall k:int. !j <= k < Arr.length (sell_order at Before) -> 0 < Uint.to_int(sell_order[k].tokens)  }

         invariant {matching !order_list (buy_order at Before) (sell_order at Before)}

         invariant {forall i:uint. 0 <= i < Arr.length (sell_order at Before) ->
                             sum_seller !order_list i + sell_order[i].tokens =  (sell_order at Before)[i].tokens  }

         invariant {forall i:uint. 0 <= i < Arr.length (buy_order at Before) ->
                             sum_buyer !order_list i + Uint.to_int(buy_order[i].tokens) =  Uint.to_int((buy_order at Before)[i].tokens)  }
         invariant { forall l. correct l (old buy_order) (old sell_order) ->
                               nb_token l <= nb_token !order_list + nb_token (!others l) }
         invariant { forall l. correct l (old buy_order) (old sell_order) ->
                               correct (!others l) buy_order sell_order }
         invariant { forall l. correct l (old buy_order) (old sell_order) ->
                               smallest_buyer_seller (!others l) !i !j
                   }

         variant {Arr.length buy_order + Arr.length sell_order - !i - !j}

         (*check if the purchase price offer is greater than or equal to the selling price*)
         if Uint.(>=) buy_order[!i].price_order sell_order[!j].price_order then begin

           (*check if the seller can provide me enough energy*)
           if Uint.(<=) buy_order[!i].tokens sell_order[!j].tokens then begin

              (*if this is the case then the quantity transferred is worth the requested quantity of the buyer*)
              let amount_transfered = buy_order[!i].tokens in

              let ghost others' = !others in
              let ghost buyer = !i in
              let ghost seller = !j in
              let ghost buy_order' : Seq.seq order = buy_order.elts in
              let ghost sell_order' : Seq.seq order = sell_order.elts in
              others := (fun l -> if pure { correct l buy_order0 sell_order0 }
                                  then remove_seller_buyer' (others' l) buy_order' sell_order' buyer seller amount_transfered
                                  else l);


              assert { forall l. correct l (old buy_order) (old sell_order) ->
                                 matching (!others l) buy_order sell_order      };

              (*I subtract from the seller the amount transferred, he can sell the energy he has in excess to another buyer*)
              sell_order[!j] <- { sell_order[!j] with tokens = Uint.(-) sell_order[!j].tokens buy_order[!i].tokens};
              buy_order[!i] <- { buy_order[!i] with tokens = 0};
              (*I have a seller a buyer and the transaction, I create a record*)
              assert { forall k: int. 0 <= k < Arr.length sell_order -> k <> !j -> sell_order[k].orderAddress == (sell_order[k].orderAddress at Before) };
              assert { forall k: int. 0 <= k < Arr.length buy_order -> k <> !i -> buy_order[k].orderAddress == (buy_order[k].orderAddress at Before) };
              
              assert { forall l. correct l (old buy_order) (old sell_order) ->
                                 matching (!others l) buy_order sell_order      };
              let registered_order = {
                  seller_index = !j;
                  buyer_index = !i;
                  amount_t = amount_transfered;
              } in
              assert { matching_order registered_order (buy_order at Before) (sell_order at Before) };

              assert { forall j: int. 0 <= j < Arr.length sell_order -> sell_order[j].orderAddress == (sell_order[j].orderAddress at Before) };

              (*I add to my list the new matching*)
              order_list := Cons registered_order !order_list;

              assert { forall l. correct l (old buy_order) (old sell_order) ->
                                 smallest_buyer_seller (!others l) !i !j
                   };

              assert { forall l. correct l (old buy_order) (old sell_order) ->
                       sum_buyer (!others l) !i = 0 };
              (*I go to the next buyer *)
              i := !i + 1;
              assert { forall l. correct l (old buy_order) (old sell_order) ->
                                 smallest_buyer_seller (!others l) !i !j
                   };
             

(*              
              assert { forall l. correct l (old buy_order) (old sell_order) ->
                               nb_token l <= nb_token !order_list + nb_token (!others l) };
              assert { forall l. correct l (old buy_order) (old sell_order) ->
                               matching (!others l) buy_order sell_order };
              assert { forall l. correct l (old buy_order) (old sell_order) ->
                               forall k :int. 0 <= k < Len.length (!others l) ->
                               !i <= (nth k (!others l)).buyer_index  /\
                               !j <= (nth k (!others l)).seller_index
              };
*)              
              (* if the seller has sold all of his energy, then I go to the next seller *)
              if sell_order[!j].tokens = 0 then begin
                  assert { forall l. correct l (old buy_order) (old sell_order) ->
                       sum_seller (!others l) !j = 0 };
                  j := !j+1;
              end
           (*if the seller does not have enough energy that the buyer wants*)
           end else begin
             (*the amount of energy sent is worth the totality of energy of the seller*)
             let amount_transfered = sell_order[!j].tokens in

              let ghost others' = !others in
              let ghost buyer = !i in
              let ghost seller = !j in
              let ghost buy_order' : Seq.seq order = buy_order.elts in
              let ghost sell_order' : Seq.seq order = sell_order.elts in
              others := (fun l -> if pure { correct l buy_order0 sell_order0 }
                                  then remove_seller_buyer' (others' l) buy_order' sell_order' buyer seller amount_transfered
                                  else l);

             (*I subtract from the buyer the amount of energy of the seller, and what remains he can buy from another seller*)
             buy_order[!i] <- { buy_order[!i] with tokens = Uint.(-) buy_order[!i].tokens sell_order[!j].tokens};
             sell_order[!j] <- { sell_order[!j] with tokens = 0 };
             assert { forall k: int. 0 <= k < Arr.length sell_order -> k <> !j -> sell_order[k].orderAddress == (sell_order[k].orderAddress at Before) };
             assert { forall k: int. 0 <= k < Arr.length buy_order -> k <> !i -> buy_order[k].orderAddress == (buy_order[k].orderAddress at Before) };
             (*I create a new record that I will store in my order list*)
              let registered_order = {
                  seller_index = !j;
                  buyer_index = !i;
                  amount_t = amount_transfered;
              } in
              order_list := Cons registered_order !order_list;
              (*I go to the next seller so that the buyer can exchange with another seller*)
              j := !j + 1
              end
           end
         else begin
           assert { forall l. correct l (old buy_order) (old sell_order) ->
                    forall k :int. 0 <= k < Len.length (!others l) ->
                      !j = (nth k (!others l)).seller_index ->
                      sell_order[!j].price_order <= buy_order[(nth k (!others l)).buyer_index].price_order
           };
           assert { sorted_order buy_order };
           j := !j + 1;  (*in case there is no matching I go to the next seller*)
         end
        done;

      (*I return my order list created*)
      !order_list
end


module Gas
  use int.Int
  use ref.Ref
  use bool.Bool

    exception Out_of_gas
    (*note that the add_gas function is different from that of the paper*)
    (*Indeed, in this version we do not take into account the allocation parameter*)
    (*the compilation and calculation of the number of gas consumed does not yet work*)
    (*on our case study, but it is in progress. So we have simplify the add_gas function.*)
    type gas = int
    val ghost tot_gas : ref gas 

    val ghost add_gas (used : gas) : unit
      requires { 0 <= used  }
      ensures  { !tot_gas = (old !tot_gas) + used }
      writes   { tot_gas }

end

module ETPMarket
  use my_library.Address
  use my_library.UInt256
  use my_library.Uint
  use my_library.SmartMeterID
  use mach.peano.Peano as Peano
  (* use my_library.PeanoUint160 as PeanoInt160 *)
  use Gas
  use int.Int
  use ref.Ref
  use Trading 

    type purchase = {amount_p: uint; price_p : uint} (*it can be buy ou sell -- amount it's the energy in tokens*)

    val marketOpen : ref bool
    constant sell_gas_consumed : gas
    constant buy_gas_consumed : gas

    axiom sell_consumed: sell_gas_consumed >= 0
    axiom buy_consumed: buy_gas_consumed >= 0

    clone my_library.Hashtbl as Ord with
          type key = Peano.t

    type ord = {
       mutable nextID: Peano.t;
       ord: Ord.t order;
    }
    invariant { 0 <= nextID }
    invariant { forall x:Peano.t. 0 <= x < nextID -> Ord.mem_ ord x }
    invariant { forall x:Peano.t. nextID <= x -> not (Ord.mem_ ord x) }
    by {
      nextID = Peano.zero;
      ord = Ord.create ();
    }

    val sellOrd : ord
    val buyOrd : ord

    exception WhenMarketOpen (*modifier WhenMarketOpen*)

    (* cf https://gitlab.inria.fr/why3/why3/merge_requests/201 *)
    axiom injectivity: forall x y: Peano.t. (x:int) = y -> x = y

      (*private function *)
      let eTPMarket_sell (_sell_purch : purchase) : unit
        requires { !marketOpen }
        requires {(_sell_purch.amount_p) > 0 }
        requires {(_sell_purch.price_p) > 0 }

        (*the function add a new order*)
        ensures { (Ord.sizee sellOrd.ord) = (Ord.sizee (old sellOrd.ord) + 1) }

        (*I found in the hashtable the sell order I recorded*)
        ensures {let order = Ord.find_ sellOrd.ord (old sellOrd.nextID) in
                 order.tokens = _sell_purch.amount_p /\
                 order.price_order = _sell_purch.price_p /\
                 order.orderAddress = msg_sender
                }

        ensures {!tot_gas - old !tot_gas <= sell_gas_consumed}
      =
        let sell_order = {
                          orderAddress = msg_sender; (*msg sender is the account address that calls this function, the seller*)
                          tokens = _sell_purch.amount_p;
                          price_order = _sell_purch.price_p;
                        } in

        Ord.add sellOrd.ord sellOrd.nextID sell_order;
        sellOrd.nextID <- Peano.succ sellOrd.nextID;
        add_gas (sell_gas_consumed)

      (*private function*)
      let eTPMarket_buy (_buy_purch : purchase) : unit
        requires { !marketOpen }
        requires { _buy_purch.amount_p > 0}
        requires { _buy_purch.price_p > 0}
        ensures { (Ord.sizee buyOrd.ord) = (Ord.sizee (old buyOrd.ord) + 1) }
        ensures {let order = Ord.find_ buyOrd.ord (old buyOrd.nextID) in
                order.orderAddress = msg_sender /\
                order.tokens = _buy_purch.amount_p /\
                order.price_order = _buy_purch.price_p
                }
        ensures {!tot_gas - old !tot_gas <= buy_gas_consumed}

      =
        let buy_order = {orderAddress = msg_sender; (*msg sender is the potential buyer who will call the buy function*)
                        tokens = _buy_purch.amount_p;
                        price_order = _buy_purch.price_p;} in
        Ord.add buyOrd.ord buyOrd.nextID  buy_order;
        buyOrd.nextID <- Peano.succ buyOrd.nextID; (*the mapping stores any purchase *)
        add_gas (buy_gas_consumed)

end

module ETPMarketBisBis
  use int.Int
  use ref.Ref
  use bool.Bool
  use my_library.Address 
  use my_library.Uint
  use ETPMarket 
  use Gas

    val algorithm : ref address
    val onlyOwner : ref bool
    val owner : address

    constant open_gas_consumed : gas
    constant close_gas_consumed : gas
    constant setAlgo_gas_consumed : gas

    axiom open_gas: open_gas_consumed >= 0
    axiom close_gas: close_gas_consumed >= 0
    axiom setAlgo_gas: setAlgo_gas_consumed >= 0

    exception OnlyOwner
    exception MarketOpen
    exception MarketClose


      (* public function *)
      let openMarket () : unit
        ensures {!tot_gas - old !tot_gas <= open_gas_consumed}
        raises {MarketOpen -> !marketOpen = True}
      =
        if !marketOpen then raise MarketOpen;
        marketOpen := True;
        add_gas (open_gas_consumed)

      (* public function *)
      let closeMarket () : unit
        ensures {!tot_gas - old !tot_gas <= close_gas_consumed}
        raises {MarketClose -> !marketOpen = False}
      =
        if not !marketOpen then raise MarketClose;
        marketOpen := False;
        sellOrd.nextID <- Peano.zero;
        Ord.clear sellOrd.ord;
        buyOrd.nextID <- Peano.zero;
        Ord.clear buyOrd.ord;
        add_gas (close_gas_consumed)

      (* public function *)
      let eTPMarket_setAlgorithm (_algoritmAddress : address)
        raises {OnlyOwner -> !onlyOwner = False} 
      =
        if not (!onlyOwner) then raise OnlyOwner;
        algorithm := _algoritmAddress;
        add_gas (setAlgo_gas_consumed)


end

module ETPAccount
  use int.Int
  use my_library.Address
  use my_library.UInt256
  use my_library.Uint
  use Gas
  use ETPMarket
  use bool.Bool
  use ref.Ref

    constant asell_gas_consumed : gas
    constant abuy_gas_consumed : gas
    constant acomplete_gas_consumed : gas

    axiom asell_gas: asell_gas_consumed >= 0
    axiom abuy_gas: abuy_gas_consumed >= 0
    axiom acomplete_gas: acomplete_gas_consumed >= 0

        (*private function*)
      let eTPAccount_sell (_sell_pursh : purchase)
        requires { !marketOpen}
        requires {(_sell_pursh.amount_p) > 0}
        requires {(_sell_pursh.price_p) > 0}
      =
        eTPMarket_sell (_sell_pursh);
        add_gas (asell_gas_consumed)


        (* private function *)
      let eTPAccount_buy (_buy_pursh : purchase)
        requires { !marketOpen}
        requires {(_buy_pursh.amount_p) > 0}
        requires {(_buy_pursh.price_p) > 0}
      =
        eTPMarket_buy (_buy_pursh);
        add_gas (abuy_gas_consumed)



        (* private function *)
      let eTPAccount_complete (_sellerAddress : address) (_callerFunction : address) (_price : uint) : unit
        requires {acceptableEtherTransaction balance  _callerFunction _sellerAddress ( _price)}
        requires {uniqueAddress _sellerAddress _callerFunction }
        requires {(_price) > 0}
        ensures {etherTransactionCompletedSuccessfully (old balance) balance _sellerAddress _callerFunction}
      =
        address_send (UInt256.v_of_uint (_price)) _callerFunction _sellerAddress;
        add_gas (acomplete_gas_consumed)
end

module ETPRegistryBis
  use my_library.UInt256
  use my_library.SmartMeterID 
  use my_library.Address
  use my_library.Uint
  use Gas
  use ETPMarketBisBis
  use ETPAccount
  use ETPMarket
  use int.EuclideanDivision
  use int.Power
  use int.Int
  use ref.Ref
  use bool.Bool
  use Trading
  use DCC

    val market : ref address
    val oracle : address
    val defAddress : address
    val onlyOracle : ref bool

    let constant floatingPointCorrection : uint = 0x10000000
    constant setMarket_gas_consumed : gas
    constant register_gas_consumed : gas
    constant record_gas_consumed : gas

    axiom setMarket_gas: setMarket_gas_consumed >= 0
    axiom register_gas: register_gas_consumed >= 0
    axiom record_gas: record_gas_consumed >= 0

    clone my_library.Hashtbl as AddressOf with
          type key = smartMeterID

    val exportBalanceOf : Bal.t uint
    val importBalanceOf : Bal.t uint
    val marketBalanceOf : Bal.t uint
    val addressOf : AddressOf.t address

    exception OnlyOracle (*modifier OnlyOracle*)
    exception OwnerNotFound
    exception ExistingSmartMeter
    exception NoSmartMeter
    exception NoAmount
    exception OverFlow
    exception ExistingRecord
    exception ExistingOrder
    exception ZeroNumber
    exception MarketNotFound
    exception ExistingMarket
    exception NoPrice

      (* public function *)
      let eTPRegistry_setMarket (_market : address)
        raises {OnlyOwner -> !onlyOwner = False}
        raises {ExistingMarket -> !market = _market}
      =
        if not !onlyOwner then raise OnlyOwner;
        if (!market == _market) then raise ExistingMarket;
        market := _market;
        add_gas (setMarket_gas_consumed)

      (* public function *)
      let registerSmartMeter (_meterID : smartMeterID) (_ownerAddress : address)
        raises { OnlyOwner-> !onlyOwner = False }
        raises {ExistingSmartMeter -> AddressOf.mem_ addressOf _meterID}
        ensures { (AddressOf.sizee addressOf) = (AddressOf.sizee (old addressOf) + 1 ) }
        ensures { AddressOf.mem_ addressOf _meterID}
      =
        if not (!onlyOwner) then raise OnlyOwner;
        if AddressOf.mem addressOf _meterID then raise ExistingSmartMeter;
        AddressOf.add addressOf _meterID _ownerAddress;
        add_gas (register_gas_consumed)

      (* public function *)
      let recordImportsAndExports (pot_buy : pot_buy) (pot_sell : pot_sell)
        raises {OnlyOracle -> !onlyOracle = False }
        raises {NoSmartMeter -> not AddressOf.mem_ addressOf pot_buy.smb_id \/ not AddressOf.mem_ addressOf pot_sell.sms_id}
        raises {OwnerNotFound -> AddressOf.([]) addressOf pot_buy.smb_id = defAddress \/ AddressOf.([]) addressOf pot_sell.sms_id = defAddress}
        raises {WhenMarketOpen -> not !marketOpen}
        raises {NoAmount -> pot_sell.amount_s = zero_unsigned \/ pot_buy.amount_b = zero_unsigned}
        raises {OverFlow -> (pot_sell.amount_s) > div (max_uint) ((floatingPointCorrection)) \/
                (pot_buy.amount_b) > div (max_uint) ((floatingPointCorrection)) \/
                (pot_sell.amount_s) * (floatingPointCorrection) > max_uint \/
                (pot_buy.amount_b) * (floatingPointCorrection) > max_uint }
        raises {ExistingRecord -> Bal.mem_ exportBalanceOf (AddressOf.([]) addressOf pot_sell.sms_id)
                \/ Bal.mem_ importBalanceOf (AddressOf.([]) addressOf pot_buy.smb_id) }
        raises {ZeroNumber -> floatingPointCorrection = zero_unsigned}
        raises {ExistingMarket -> Bal.mem_ marketBalanceOf !market}
        raises {NoPrice -> pot_sell.price_s <= 0 \/ pot_buy.price_b <= 0}
      =
        if not !marketOpen then raise WhenMarketOpen;
        if not (!onlyOracle) then raise OnlyOracle;
        if not AddressOf.mem addressOf pot_buy.smb_id then raise NoSmartMeter;
        if not AddressOf.mem addressOf pot_sell.sms_id then raise NoSmartMeter;

        let owner_s = AddressOf.find_def addressOf pot_sell.sms_id defAddress in
        if owner_s == defAddress then raise OwnerNotFound;
        
        let owner_b = AddressOf.find_def addressOf pot_buy.smb_id defAddress in
        if owner_b == defAddress then raise OwnerNotFound;
        if pot_buy.amount_b = 0 then raise NoAmount;
        if pot_sell.amount_s = 0 then raise NoAmount;
        if floatingPointCorrection = 0 then raise ZeroNumber;
        if (pot_sell.amount_s) > (Uint.(/) (Uint.of_int(max_uint)) floatingPointCorrection) then raise OverFlow;
        if (pot_buy.amount_b) > (Uint.(/) (Uint.of_int(max_uint)) floatingPointCorrection) then raise OverFlow;
        let exportWithCorrection = (pot_sell.amount_s) * (floatingPointCorrection) in
        if Bal.mem exportBalanceOf owner_s then raise ExistingRecord;
        if Bal.mem importBalanceOf owner_b then raise ExistingRecord;
        if pot_sell.price_s <= 0 then raise NoPrice;
        if pot_buy.price_b <= 0 then raise NoPrice;

        let export_purchase = {
                                amount_p = exportWithCorrection;
                                price_p = pot_sell.price_s;
                               } in
        Bal.add exportBalanceOf owner_s ((export_purchase).amount_p);

        let importWithCorrection =  (pot_buy.amount_b) * (floatingPointCorrection) in
        let import_purchase = {
                                amount_p = importWithCorrection;
                                price_p = pot_buy.price_b;
                              } in

        Bal.add importBalanceOf owner_b ((import_purchase).amount_p);

        if Bal.mem marketBalanceOf !market then raise ExistingMarket;
        Bal.add marketBalanceOf !market 0;
        if (pot_buy.amount_b > 0) then eTPAccount_buy(import_purchase)
        else eTPAccount_sell(export_purchase);
        add_gas (record_gas_consumed)

end

module ETPRegistry
  use int.Int
  use my_library.UInt256
  use my_library.SmartMeterID
  use my_library.Address
  use my_library.Uint
  use ref.Ref
  use ETPMarket
  use Gas
  use ETPRegistryBis
  use bool.Bool


  val onlymarket : ref bool (*modifier*)
  constant transferTo_gas_consumed : gas
  constant transferFrom_gas_consumed : gas

  axiom transferTo_gas: transferTo_gas_consumed >= 0
  axiom transferFrom_gas: transferFrom_gas_consumed >= 0

    (* private function *)
    let transferToMarket (_from : address) (_value : uint) : unit (* value are green tokens to send *)
      
      requires {!onlymarket}
      requires { _value > 0 }
      requires { (Bal.([]) marketBalanceOf !market) = 0 }
      requires { acceptableAmountTransaction exportBalanceOf marketBalanceOf _from !market _value}
      ensures {amountTransactionCompletedSuccessfully (old exportBalanceOf) exportBalanceOf (old marketBalanceOf) marketBalanceOf _from !market }
    =
      amount_transaction (exportBalanceOf) (marketBalanceOf) (_from) (!market) (_value);
      add_gas (transferTo_gas_consumed)

    (* private function *)
    let transferFromMarket (_to : address) (_value : uint) : unit (*_value = green token*)
      
      requires {!onlymarket}
      requires {_value > 0 }
      requires {(Bal.([]) marketBalanceOf !market) > 0}
      requires {acceptableAmountTransaction marketBalanceOf importBalanceOf !market _to _value}
      ensures {amountTransactionCompletedSuccessfully (old marketBalanceOf) marketBalanceOf (old importBalanceOf) importBalanceOf !market _to}
      
    =
      amount_transaction (marketBalanceOf) (importBalanceOf) (!market) (_to) (_value);
      add_gas (transferFrom_gas_consumed)

end

module ETPMarketBis
  use int.Int
  use my_library.SmartMeterID
  use my_library.Address
  use my_library.UInt256
  use my_library.Uint
  use Gas
  use ETPMarket
  use ETPAccount
  use ETPRegistry
  use ETPRegistryBis
  use ref.Ref
  use Trading



    
    val onlyAlgo : ref bool (*modifier*)
    constant mcomplete_gas_consumed : gas
  
    axiom mcomplete_gas: mcomplete_gas_consumed >= 0

      (* private function *)
      let eTPMarket_complete (sellId: Peano.t) (buyId : Peano.t) (_purchase : purchase) : unit
        requires {!onlymarket}
        requires { (_purchase.amount_p) > 0 /\  (_purchase.price_p) > 0 }
        requires {(Bal.([]) marketBalanceOf !market) > 0}
        requires {acceptableAmountTransaction marketBalanceOf importBalanceOf !market ((Ord.([]) buyOrd.ord buyId).orderAddress) _purchase.amount_p}
        requires {acceptableEtherTransaction balance (Ord.([]) buyOrd.ord buyId).orderAddress (Ord.([]) sellOrd.ord sellId).orderAddress ( _purchase.price_p)}
        
        requires {!onlyAlgo}
        requires { sellId >= 0 /\ buyId >= 0 }
        requires {Ord.mem_ sellOrd.ord sellId}
        requires {Ord.mem_ buyOrd.ord buyId}
        
        requires {uniqueAddress (Ord.([]) sellOrd.ord sellId).orderAddress (Ord.([]) buyOrd.ord buyId).orderAddress}

      
        ensures {etherTransactionCompletedSuccessfully (old balance) balance (Ord.([]) buyOrd.ord buyId).orderAddress (Ord.([]) sellOrd.ord sellId).orderAddress}
        ensures {amountTransactionCompletedSuccessfully (old importBalanceOf) importBalanceOf (old marketBalanceOf) marketBalanceOf (Ord.([]) buyOrd.ord buyId).orderAddress !market}
      =
        let sellOrder = Ord.([]) sellOrd.ord sellId in
        let buyOrder = Ord.([]) buyOrd.ord buyId in
        eTPAccount_complete (sellOrder.orderAddress) (buyOrder.orderAddress) (_purchase.price_p);
        transferFromMarket (buyOrder.orderAddress) (_purchase.amount_p);
        add_gas (mcomplete_gas_consumed)
end



\end{lstlisting}

\section{: WCET of function with allocation}

\centering

\begin{lstlisting}[language = Why3, basicstyle=\fontsize{8}{10}\tt]
type list 'a = Nil | Cons 'a (list 'a)

function length (l: list 'a) : int =
  match l with
  | Nil      -> 0
  | Cons _ r -> 1 + length r
  end

let rec length_ [@ evm:gas_checking] (l:list 'a) : int32
 requires { (length l) <= max_int32 }
 ensures { !gas - old !gas <= (length l) * 128 + 71 }
 ensures { !alloc - old !alloc <= 0 }
 ensures { result = length l }
 variant { l } =
  match l with
  | Nil -> add_gas 71 0; 0
  | Cons _ l -> add_gas 128 0; 1 + length_ l
  end

let rec mk_list42 [@ evm:gas_checking] (i:int32) : list int32
 requires { 0 <= i }
 ensures { !gas - old !gas <= i * 185 + 113 }
 ensures { !alloc - old !alloc <= i * 96 + 32 }
 ensures { i = length result }
 variant { i } =
  if i <= 0 then (add_gas 113 32; Nil) else
  let l = mk_list42 (i-1) in
  add_gas 185 96;
  Cons (0x42:int32) l

let g_ [@ evm:gas_checking] (i:int32) : int32
 requires { 0 <= i }
 ensures { !gas - old !gas <= i * 313 + 242 }
 ensures { !alloc - old !alloc <= i * 96 + 32 } =
  add_gas 58 0;
  let l = mk_list42 i in
  length_ l
\end{lstlisting}

\end{subappendices}


% \bibitem{nakamoto}
% Nakamoto, S.: Bitcoin: A peer-to-peer electronic cash system. Working Paper, (2008)




% LNCS Homepage, \url{http://www.springer.com/lncs}. Last accessed 4
% \end{thebibliography}
\end{document}